\documentclass[preprint,showpacs,preprintnumbers,amsmath,amssymb,superscriptaddress,floatfix]{revtex4}
\usepackage{graphicx}

\begin{document}
\title{Measurement method for the nuclear anapole moment of laser trapped alkali atoms}
\author{E. Gomez\footnote{Present address: National Institute of Standards and Technology, Maryland, USA.}, S. Aubin\footnote{Present address:
Department of Physics, University of Toronto, Ontario, Canada.}, and
G. D. Sprouse} \affiliation{Dept. of Physics and Astronomy, Stony
Brook University, Stony Brook, NY 11794-3800}
\author{L. A. Orozco}
\affiliation{Dept. of Physics, University of Maryland, College
Park, MD 20742-4111}
\author{D. P. DeMille}
\affiliation{Dept. of Physics, Yale University, New Haven, CT
06520-8120}
\date{\today}

\begin{abstract}
Weak interactions within a nucleus generate a nuclear spin
dependent, parity violating electromagnetic moment, the anapole
moment. We analyze a method to measure the nuclear anapole moment
through the electric dipole transition it induces between hyperfine
states of the ground level. The method requires tight confinement of
the atoms to position them at the anti-node of a standing wave Fabry
Perot cavity driving the anapole-induced micro-wave E1 transition.
We explore the necessary limits in the number of atoms, excitation
fields, trap type, interrogation method, and systematic tests
necessary for such measurements in francium, the heaviest alkali.
\end{abstract}

\maketitle

\section{Introduction}

Zel'dovich postulated in 1957 that the weak interactions between
nucleons would generate a parity violating, time reversal conserving
moment called the anapole moment~\cite{zeldovich58}. Flambaum and
Khriplovich calculated the effect it would have in
atoms~\cite{flambaum84}. Experiments in thallium gave a limit for
its value~\cite{vetter95}, and it was measured for the first time
with an accuracy of 14\% through the hyperfine dependence of atomic
parity nonconservation (PNC) in cesium~\cite{wood99,wood97}.

We present in this paper a measurement strategy of the nuclear
anapole moment by direct excitation of the microwave electric dipole
(E1) transition between the ground hyperfine levels in a chain of
isotopes of an alkali atom. Alkali atoms are the best understood
quantitatively in their electronic properties associates with PNC.
The precision of the Cs PNC experiments has required more detailed
studies of the nuclear structure~\cite{derevianko02}. Measurements
over a chain of isotopes offer the advantage that they can focus on
the differences appearing as the number of neutrons changes. This
task has been accomplished well by theory (see for example
Ref.~\cite{grossman99}) for the hyperfine anomaly measurements in
Fr.

Current plans at the Isotope Separator and Accelerator (ISAC) at
TRIUMF, in Vancouver Canada, should provide access to all the
neutron deficient long lived isotopes of Fr with lifetimes above 30
s and to a similar number of the neutron rich isotopes, a sufficient
variety to give a difference in number of neutrons of more than 10.
The expected production rates should be at least two orders of
magnitude larger than those obtained at Stony Brook, the leading
place for study of Fr \cite{gomez06}.

A measurement of the anapole moment in a chain of isotopes will
provide information about neutral weak currents in the nucleus. The
measurements can also give information on the nuclear structure and
its changes as the number of neutrons increases
\cite{fortson90,pollock92}.

The E1 transition between hyperfine levels is parity forbidden, but
becomes allowed by the anapole induced mixing of levels of opposite
parity. The general approach has been suggested in the
past~\cite{loving77,gorshkov88,budker,balakin80,novikov75,hinds77,adelberger81,fortson93}.
We would place many atoms inside a microwave Fabry-Perot cavity and
hold them in a blue detuned dipole trap. The atoms would interact
with the microwave field and with a Raman field generated by a pair
of laser beams, in the presence of a static magnetic field. We would
confine the atoms to the node (anti-node) of the magnetic (electric)
microwave field to drive only an E1 transition between hyperfine
levels. The atoms would start in the lower hyperfine level, with the
signal proportional to the population of atoms in the upper
hyperfine level after the excitation. The interference with a Raman
transition would give a signal linear in the E1 transition.

Recent work related to time-reversal invariance tests in atomic
traps \cite{romalis99,chin01}, points to the many potential
advantages of combining traps with tests of fundamental symmetries,
but also highlights the potential systematic errors present in such
measurements, making a careful evaluation of the method prior to its
implementation necessary. We focus our study primarily on isotopes
of francium, the heaviest of the alkali atoms~\cite{gomez06}, in an
optical dipole trap, where the effect is expected to be large.

The organization of the paper is as follows: Section II gives the
theoretical background for the nuclear anapole moment, section III
explains the proposed measurement method, section IV presents an
analysis of noise sources and systematic effects, and section V
contains the conclusions.

\section{Theoretical background}

The exchange of weak neutral currents between electrons and nucleons
constitute the main source of parity violating atomic transitions.
The currents are of two kinds, depending on whether the electron or
the nucleon enters as the axial vector current. The corresponding
terms in the Hamiltonian differ on their dependence on the nuclear
spin. The part independent of the nuclear spin is generally the
dominant contribution in atomic PNC. This is not the case for the
present work, where we consider transitions between hyperfine levels
of the ground state, and the contribution from the nuclear spin
independent part is zero~\cite{angstmann05}. The Hamiltonian for the
spin dependent part in the shell model with a single valence nucleon
of unpaired spin is given by \cite{flambaum97}

\begin{equation}
H=\frac{G}{\sqrt{2}} \frac{K \mbox{\boldmath$I \cdot
\alpha$}}{I(I+1)} \kappa_{i} \delta(r), \label{hpnc}
\end{equation}
where $G =10^{-5}$ m$_p^{-2}$ is the Fermi constant, m$_p$ is the
proton mass, $K=(I+1/2)(-1)^{I+1/2-l}$, $l$ is the nucleon orbital
angular momentum, $I$ is the nuclear spin, ${\boldmath \alpha}$ are
Dirac matrices, and $\kappa_{i}$ is the interaction constant, with
$i=p,n$ for a proton or a neutron. The terms proportional to the
anomalous magnetic moment of the nucleons and the electrons have
been neglected.

The interaction constant is given by \cite{flambaum97}

\begin{equation}
\kappa_{i}=\kappa_{a,i}- \frac{K-1/2}{K}\kappa_{2,i}+
\frac{I+1}{K}\kappa_{Q_W}, \label{kappatotal}
\end{equation}
with $\kappa_{2,p}=-\kappa_{2,n}=-1.25(1-4\sin^2\theta_W)/2$,
corresponding to the tree level approximation, with $\sin^2\theta_W
\sim 0.23$ the Weinberg angle. Equation \ref{kappatotal} has two
corrections, $\kappa_{a,i}$ the effective constant of the anapole
moment, and $\kappa_{Q_W}$ that is generated by the nuclear spin
independent part of the electron nucleon interaction together with
the hyperfine interaction. The three parts of this interaction
constant can be traced to different ways in which the weak
interacting vector boson $Z^0$ appears in the Feynman diagrams. The
first one (the anapole) correspond to vertex corrections due to weak
hadronic interactions on the nuclear side of the electromagnetic
interaction coupled to the electron through a virtual photon. The
second one takes the direct effect of a $Z^0$ exchange between the
electron vector current and the nuclear axial current. The last one
is the simultaneous exchange of a $Z^0$ and a photon modifying the
hyperfine interaction. Flambaum and Murray showed that
\cite{flambaum97}

\begin{eqnarray}
& & \kappa_{a,i}= \frac{9}{10} g_i \mu_i \frac{\alpha {\cal
A}^{2/3}}{m_p \tilde{r_0}}, \nonumber
\\ & & \kappa_{Q_W}=-\frac{1}{3} \left(\frac{Q_W}{{\cal A}}\right) \mu_N \frac{\alpha {\cal A}^{2/3}}{m_p \tilde{r_0}},
\label{kappaa}
\end{eqnarray}
where $\alpha$ is the fine structure constant, $\mu_i$ and $\mu_N$
are the magnetic moments of the external nucleon and of the nucleus
respectively, $\tilde{r_0}=1.2~$fm, ${\cal A}$ is the atomic mass
number, $Q_W$ is the weak charge, and $g_i$ gives the strength of
the weak nucleon-nucleus potential with $g_p \sim 4$ for a proton
and $0.2<g_n<1$ for a neutron~\cite{khriplovich91}. The anapole
moment is the dominant contribution to the interaction in heavy
atoms, for example, $\kappa_{a,p}$/$\kappa_{Q_W} \simeq$15 for
$^{209}$Fr. We will assume from now on that
$\kappa_{i}=\kappa_{a,i}$.

\subsection{The anapole moment}

The anapole moment of a nucleus is a parity non-conserving (PNC),
time reversal conserving moment that arises from weak interactions
between the nucleons (see the review by Haxton and
Wieman~\cite{haxton01}.) It can be detected in a PNC
electron-nucleus interaction, and reveals itself in the spin
dependent part of the PNC interaction. Wood {\it et al.}
\cite{wood99,wood97} measured the anapole moment of $^{133}$Cs by
extracting the dependence of atomic PNC on the hyperfine levels
involved.

The anapole moment classically is defined by (see for example
\cite{lewis93})

\begin{equation}
{\bf a}=-\pi \int d^3r r^2 {\bf J(r)}, \label{definea}
\end{equation}
with ${\bf J}$ the electromagnetic current
density~\cite{noteanapole}. The nuclear anapole moment in francium
arises mainly from the weak interaction between the valence nucleons
and the core. Flambaum, Khriplovich and Sushkov \cite{flambaum84}
estimate the anapole moment from Eq. \ref{definea} for a single
valence nucleon to be
\begin{equation}
{\bf a} =\frac{G}{e\sqrt{2}}\frac{K}{j(j+1)}\kappa_{a,i}~
\mbox{\boldmath$j$} = C^{an}_i \mbox{\boldmath$j$},
\label{anapolemoment}
\end{equation}
where $j$ is the nucleon angular momentum and $e$ the electron
charge. The calculation assumes a homogeneous nuclear density, and a
core with zero angular momentum, leaving the valence nucleon
carrying all the angular momentum.

The measurement of the anapole moment gives information on the
weak nucleon-nucleon interactions. A measurement of the anapole
moment in a chain of isotopes would provide a separation of the
anapole moment due to the valence proton or neutron.

\subsection{Calculations of the anapole moment of francium isotopes}

\begin{figure}
\leavevmode \centering
\includegraphics[width=3.1in]{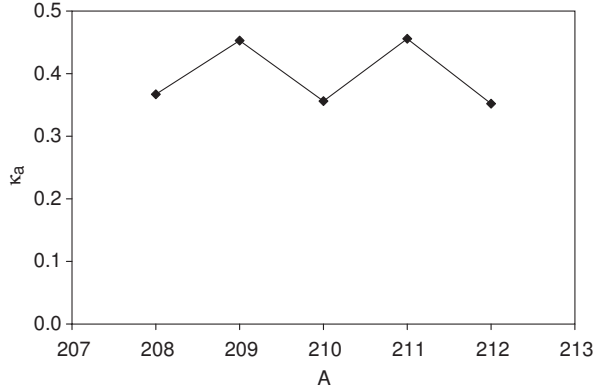}
\caption{Anapole moment effective constant for different isotopes
of francium. \label{isotopes}}
\end{figure}

We use Eqs. \ref{kappaa} and \ref{anapolemoment} to estimate the
anapole moment of five light francium isotopes with radioactive
lifetimes longer than one minute~\cite{grossman99}. The unpaired
valence proton generates the anapole moment in even-neutron
isotopes, whereas in the odd-neutron isotopes both the unpaired
valence proton and neutron participate. Francium has an unpaired
$h_{9/2}$ proton for all the isotopes and a $f_{5/2}$ neutron for
the odd-neutron isotopes around $^{210}$Fr. The protonic and
neutronic contributions add vectorially to generate the anapole
moment:

\begin{equation}
{\bf a} =\frac{C^{an}_p \mbox{\boldmath$j$}_p\cdot
\mbox{\boldmath$I$}+C^{an}_n \mbox{\boldmath$j$}_n\cdot
\mbox{\boldmath$I$}}{\mbox{\boldmath$I$}^2} \mbox{\boldmath$I$}
=\frac{G}{e\sqrt{2}}\frac{(I+1/2)}{I(I+1)}\kappa_{a}~
\mbox{\boldmath$I$}, \label{anapolesum}
\end{equation}
with $C^{an}_i \mbox{\boldmath$j$}_i$ the anapole moment for a
single valence nucleon $i$ (proton or neutron) as given by Eq.
\ref{anapolemoment} ($j_p=9/2$, $j_n=5/2$). Equation
\ref{anapolesum} defines the coupling strength of the total
anapole moment ($\kappa_a$) resulting from adding the valence
proton and neutron. Figure \ref{isotopes} shows the predicted
values of $\kappa_a$ for a string of francium isotopes
\cite{grossman99} using $g_n=1$.

\subsection{Perturbation theory} \label{perturbationtheorysection}

The anapole moment induces a small mixing of electronic states of
opposite parity. The effect of the anapole moment Hamiltonian on
the ground state hyperfine levels according to first order
perturbation theory is

\begin{equation}
|\overline{sFm}\rangle= |sFm\rangle +\sum_{F'm'}\frac{\langle
pF'm'|{\bf H_a} |sFm\rangle }{E_p-E_s}|pF'm'\rangle,
\label{perturbation}
\end{equation}
where $E_p$, and $E_s$ are the energies of the $p$, and $s$ states
respectively, and

\begin{equation}
{\bf H_a}= |e| \mbox{\boldmath$ \alpha \cdot$}{\bf{a}} \delta
({\bf r}), \label{anapolehamiltonian}
\end{equation}
is the anapole moment Hamiltonian from Eq. \ref{hpnc}, with $\bf{a}$
the anapole moment from Eq. \ref{anapolemoment}. The matrix element
in Eq. \ref{perturbation} gives \cite{khriplovich91}
\begin{eqnarray}
& \langle pF'm'| & {\bf H_a} ~|sFm\rangle = i \frac{\xi {\rm Z}^2
R}{(\varrho_s \varrho_p)^{3/2}} \frac{2 \gamma +1}{3} \frac{(I+1/2)
\kappa_a Ry}{I(I+1)} \nonumber
\\ & & \times (F(F+1)-I(I+1)-3/4) \delta_{F,F'} \delta_{m,m'},
\label{perturbation2}
\end{eqnarray}
with $\xi =G m_e^2 \alpha^2/ \sqrt{2}\pi=3.651 \times 10^{-17}$,
$m_e$ the electron mass, $\rm Z$ the atomic number, $\varrho_s$ and
$\varrho_p$ the effective principal quantum number for the $s$ and
$p$ electronic states, $\gamma=\sqrt{(J+1/2)^2-{\rm Z}^2\alpha^2}$,
$J$ the electron total angular momentum, and $Ry$ the Rydberg. The
relativistic enhancement factor R is given by
\begin{eqnarray}
R=4(a_0/2{\rm Z}r_0)^{2-2\gamma}/\Gamma^2(2\gamma+1),
\label{R_factor}
\end{eqnarray}
with $a_0$ the Bohr radius, and $r_0=\tilde{r_0}{\cal A}^{1/3}$.

The anapole moment mixes only states with the same $F$ and $m$, and
the mixing grows as ${\rm Z}^{8/3}R$. For the $^{209}$Fr ground
state, we obtain

\begin{eqnarray}
& |\overline{sFm}\rangle =|sFm\rangle & - ~i~5.9\times 10^{-13}
\kappa_a \nonumber
\\ & & \times (F(F+1)-25.5) |pFm\rangle. \label{perturbation3}
\end{eqnarray}
The mixing coefficient is imaginary due to time reversal symmetry.
In practice, the mixing would be measured through the $E1$
transition amplitude $A_{E1}$ (Eq.~\ref{e1}) it induces between two
hyperfine levels. The effect in francium is 11 times larger than in
cesium~\cite{johnson03}.

\section{Proposed Measurement Strategy}
\label{measurementstrategy}

High efficiency magneto-optical traps (MOT) for francium atoms on
line with an accelerator have been demonstrated~\cite{aubin03a}.
Their performance and reliability matches the needs of the current
proposed measurement strategy. Atoms captured on a first trap would
then be transferred to a second MOT in a separated chamber. We would
load the atoms into a dipole trap located at the electric field
anti-node of a standing wave in a microwave Fabry-Perot cavity. We
would optically pump them into a single Zeeman sublevel, and prepare
a coherent superposition of the hyperfine ground levels with a Raman
pulse of amplitude $A_R$ and duration $t_R$. Simultaneously we would
drive the E1 transition of amplitude $A_{E1}$ with the cavity
microwave field, and measure the population in the upper hyperfine
level (normalized to the total number of atoms [$N$]) using a
cycling transition. The population in the upper hyperfine level at
the end of each sequence would be

\begin{equation}
\Xi_{\pm}=N|c_e|^2=N\sin^2 \left(\frac{(A_R \pm A_{E1})t_R}{2\hbar}
\right), \label{excited}
\end{equation}
where $c_e$ is the upper hyperfine level amplitude. The sign depends
on the handedness of the coordinate system defined by the external
fields, as explained in the next section. The signal for the
measurement:

\begin{eqnarray}
{\cal S} = \Xi_+- \Xi_- & & =N\sin\left(\frac{A_R t_R}{\hbar}\right)
\sin\left(\frac{A_{E1} t_R}{\hbar}\right) \nonumber \\ & & \simeq
N\sin\left(\frac{A_R t_R}{\hbar}\right) \left(\frac{A_{E1}
t_R}{\hbar}\right), \label{asymmetry}
\end{eqnarray}
would be the difference between populations in the upper hyperfine
level for both handedness. The last step assumes a small $A_{E1}$,
the quantity proportional to the anapole moment constant
$\kappa_{a}$.

\subsection{Apparatus setup}

\begin{figure}
\leavevmode \centering
\includegraphics[width=3.1in]{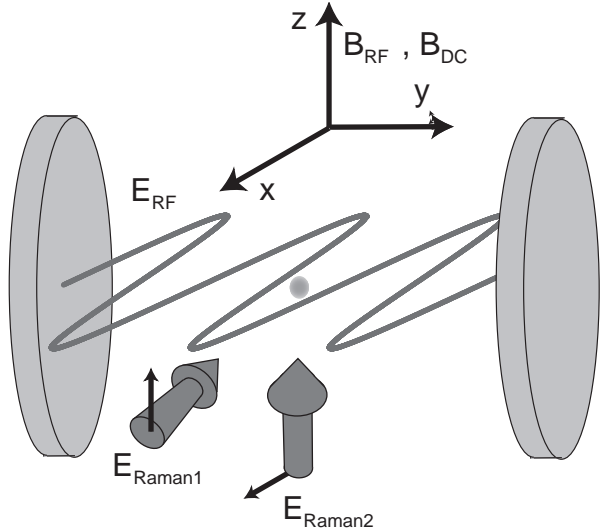}
\caption{Schematic setup of the proposed apparatus. The microwave
cavity axis is along the $y$-axis. The microwave electric field
inside the cavity oscillates along the $x$-axis. The two Raman laser
beams are polarized along the $x$-axis and $z$-axis, respectively.
The microwave magnetic field and the static magnetic field are both
directed along the $z$-axis. A dipole trap (not shown) holds the
atoms at the origin that coincides with an anti-node of the
microwave electric field. \label{setup}}
\end{figure}

Figure \ref{setup} shows a diagram of the proposed apparatus. The
atoms would be placed inside a microwave Fabry-Perot cavity at the
electric field anti-node, confined in a blue detuned dipole trap to
a volume with $10~\mu$m length along the cavity axis, and a $1$ mm
diameter in the radial dimension. Observation of the electric dipole
(E1) microwave transitions would be done through an interference
method and extraction of the signal would require repeating the
excitation varying the coordinate system.

Preparation of the atoms in a particular Zeeman sublevel of the
lower hyperfine level $|F_1,m_1\rangle$ in an applied static
magnetic field ${\bf B}=B_0 {\bf \hat{z}}$ would be necessary. A
resonant standing-wave microwave electric field ${\bf E}(t)=E
\cos(2\pi \nu_m t + \psi) \cos(k_m y) {\bf \hat{x}}$ would excite
the atoms to a particular Zeeman sublevel in the upper hyperfine
level $|F_2,m_2\rangle$. The microwave magnetic field ${\bf M}$
would be aligned along ${\bf B}$, and it is $\pi$/2 out of phase
(for a perfect standing wave) with ${\bf E}$ so that ${\bf M}(t)=M
\sin(2\pi \nu_m t + \psi) \sin(k_m y) {\bf \hat{z}}$, with $M=E$ in
cgs units.

The Raman transition would include two plane-wave optical fields,
${\bf E}_{R1}(t)=E_{R1} \cos(\omega_{R} t + \phi_{R}) {\bf \hat{x}}$
and ${\bf E}_{R2}(t)=E_{R2} \cos((\omega_{R} + 2\pi \nu_m)t +
\phi_{R}) {\bf \hat{z}}$, phase locked to the microwave field. The
Raman carrier frequency $\omega_{R}$ would be tuned sufficiently far
from optical resonance that only the vector part of the Raman
transition amplitude (${\bf V}$ $\propto i {\bf E}_{R1} \times {\bf
E}_{R2}$) would be non-negligible~\cite{demille98}; that is, we
ignore the tensor part of the Raman amplitude.

\subsection{Observable and reversals}

The various electric and magnetic fields of the apparatus would
define a coordinate system related to the measured rate $\Xi_{\pm}$.
The transition rate $\Xi_{\pm}$ depends on three vectors: The
polarization of the E1 transition ({\bf E}), the polarization of the
Raman transition (${\bf V}$), and the static magnetic field ${\bf
B}$ that provides an axis for the spins of the nuclei. We combine
these three vectors to produce the time reversal preserving pseudo
scalar $i{\bf (E \times (E}_{R1} {\bf \times E}_{R2} {\bf ) \cdot B
)}$, proportional to the measured quantity.

A single reversal of any of the fields in the above pseudo scalar
changes the sign of the interference term of $\Xi_{\pm}$. We then
would have the following reversals:

1. - Magnetic field reversal ($\beta$ reversal).

2. - A shift of $\pi$ in the relative phase between the E1 and the
Raman fields ($s$ reversal).

The Zeeman sublevels reverse with the magnetic field. The state
preparation has to be inverted in order to reach the correct
Zeeman sublevel, meaning that $\sigma^+$ light goes into
$\sigma^-$ and vice versa. The magnitude of the static magnetic
field and the microwave cavity frequency remain unchanged for this
reversal.

\subsection{Apparatus requirements}

\subsubsection{Magnetic field}

We would drive E1 transitions between two particular Zeeman
sublevels, $|F_1, m_1\rangle \rightarrow |F_2, m_2\rangle$ in
different hyperfine levels of the ground state. While the
frequencies of the exciting fields can be well controlled, the
energy difference of the Zeeman states is determined primarily by
the static magnetic field.

The experimental design should minimize the sensitivity to magnetic
field fluctuations. The energy difference between two levels passes
through a minimum at the static magnetic field $B_0$, and depends
quadratically on the magnetic field around that point. We would use
the Zeeman sublevels that give the smallest quadratic dependence.
Table \ref{tableparameters} lists the Zeeman sublevels and magnetic
fields selected for different francium isotopes. The experiment
would work between the $|F_1, m_1\rangle$ and $|F_2, m_2\rangle$
levels and also between the $|F_1, m_2\rangle$ and $|F_2,
m_1\rangle$ levels, interchanging $m_1$ and $m_2$. The operating
point of the static magnetic field and the frequency of the
microwave cavity would have to be corrected slightly because of the
nuclear spin contribution. The state preparation would also change
to start in the appropriate level. The change of $m_1$ ($m_2$) for
$m_2$ ($m_1$) does not work as a reversal because of the difference
in transition amplitude, but it can still be useful as a consistency
check.

\begin{table}
\leavevmode \centering \caption{Parameters of the five relevant
francium isotopes: Spin, hyperfine splitting (Hfs) of the $7s_{1/2}$
state~\protect\cite{coc85,coc87}, Zeeman sublevels $\rm{m_1}$,
$\rm{m_2}$, and their energy separation $\nu_m$ at the static
magnetic field $B_0$ used in the proposed measurement.}
\begin{tabular} {ccccccc}
Isotope & Spin & Hfs(MHz) & $\rm{m_1}$ & $\rm{m_2}$ & $B_0$(Gauss) & $\nu_m$(Mhz) \\
\hline
208     & 7    & 49880.3  & 0.5        & 1.5        & 2386.5       & 49433        \\
209     & 9/2  & 43033.5  & 0          & -1         & 1553.0       & 42816        \\
210     & 6    & 46768.2  & 0.5        & 1.5        & 2586.4       & 46208        \\
211     & 9/2  & 43569.5  & 0          & -1         & 1572.3       & 43349        \\
212     & 5    & 49853.1  & 0.5        & 1.5        & 3265.7       & 49015 %
\end{tabular}
\label{tableparameters}
\end{table}

The frequency for the $F=4$, $m=0$ to the $F=5$, $m=-1$ transition
in $^{209}$Fr, expanded around the critical field $B_{0}=1553$
Gauss, is

\begin{equation}
\nu_m = 42.816\times 10^9+90(B-B_0)^2 \rm{Hz},
\label{magneticfield}
\end{equation}
with $B$ in Gauss. Control of the magnetic field to 0.06 Gauss
(three parts in $10^5$) reduces the frequency noise due to
magnetic field fluctuations down to $\Delta \nu_m \sim 0.3$ Hz.

The experiment would take place in a large magnetic field whereas
the state preparation and detection occur in a small magnetic field.
The transition between both regimes should be done adiabatically.
The time scale is determined by the precession time in a small
magnetic field, resulting in a magnetic field ramp duration of
hundreds of microseconds.

\subsubsection{The microwave cavity}

The francium hyperfine separation requires a Fabry-Perot microwave
cavity operating at around 45 GHz (wavelength $\lambda_m \sim 0.66$
cm) in a Fabry-Perot configuration; for example a cavity with a
mirror separation of $d \sim 20 \lambda_m \sim 13$ cm and a mirror
radius of $r_m=3.5$ cm. These parameters combine to minimize
diffraction losses as the Fresnel number $F_N>1$, where $F_N = r_m^2
/ \lambda_m d$~\cite{ramo93}.

The quality factor ($Q$) of the cavity is
\begin{equation}
Q=\frac d{2\varsigma}, \label{q}
\end{equation}
where $\varsigma$ is the skin depth and is equal to $\sqrt{2/\omega
\mu_0 \sigma}$ with $\mu_0$ the magnetic constant and $\sigma$ the
conductivity ($5.8 \times 10^7~\Omega^{-1}$m$^{-1}$ for copper at
room temperature). The conductivity limited quality factor is $Q=1.9
\times 10^5$. It is possible to couple 58 mW into the cavity with
current available technology, which would give an electric field of
476 V/cm to drive the E1 transition.

The E1 transition amplitude for $^{209}$Fr between the initial
hyperfine level ($\overline{i}$) $F=4$, $m=0$ to the final hyperfine
level ($\overline{f}$) $F=5$, $m=-1$ with a static magnetic field of
$1553$ Gauss (see Table \ref{tableparameters}) is

\begin{eqnarray}
A_{E1}/\hbar & & = \langle \overline{f}|-e {\bf E \cdot r}
|\overline{i}\rangle/\hbar \nonumber
\\ & & =0.01i \left[\frac{E}{476 {\rm V/cm}}\right]
\left[\frac{\kappa_a}{0.45}\right] ~~{\rm rad/s}. \label{e1}
\end{eqnarray}
A more accurate result can be obtained with the use of many-body
perturbation theory~\cite{johnson03,porsev01,bouchiat91}.

A 1 cm cavity waist would cover the atoms in the 1 mm diameter 10
$\mu$m length trap, and radius of curvature of $R_m=9.9$ cm for the
cavity mirrors ensure a stable cavity, since $( 1 - ( d / 2R_m )
)^2< 1$. The curvature of the wave fronts could create a gradient of
polarization of the microwave field smaller than $3 \times 10^{-5}$
rad cm$^{-1}$ over the volume of the trap. We show later that this
rotation is within acceptable ranges.

The field inside the cavity can be decomposed into a standing wave
and a travelling wave. The presence of the travelling wave generates
M1 transitions despite the location of the atoms at the node of the
standing wave magnetic field. Significant reduction of the amplitude
of intra-cavity travelling waves comes with a symmetrical
arrangement of identical antennas, one on each mirror. Antennas give
a high coupling efficiency into the cavity \cite{harbarth87} as
compared to a slit or a grating~\cite{dees65}. The electric field
inside the cavity is given by

\begin{eqnarray}
E=e^{-i\nu_m t} & & \left(\frac{1}{1-r_1r_2 e^{2ik_m d}}\right)
\nonumber \\
& & \times \left[E_1t_1\left(e^{ik_m z}-r_2e^{ik_m d}e^{-ik_m
z}\right)
\right. \nonumber \\
& & \left. +E_2t_2\left(e^{-ik_m z}-r_1e^{ik_m d}e^{ik_m
z}\right)\right], \label{cavityfield}
\end{eqnarray}
where $r$ is the reflectivity, $t$ the transmissivity, $k$ is the
wave-vector of the microwave field, $d$ the separation between the
mirrors, and the sub indices 1 and 2 refer to the two mirrors. The
first (second) term is the field generated by antenna 1 (2). The
expression is the sum of two waves, one travelling to the right and
the other to the left. The difference in amplitude between these two
contributions results in a travelling wave. The ratio of travelling
to standing wave assuming a symmetrical cavity, that is $r_1=r_2=r$
and $t_1=t_2=t$, is

\begin{equation}
{\cal R}_{T/S} = \left(\frac{i\vartheta}{4}
+\frac{E_1-E_2}{4E_1}\right) \left(i(1-r)+k_m\Delta d\right),
\label{standingwave}
\end{equation}
with $\vartheta$ the phase mismatch from both antennas and $\Delta
d$ the deviation of the cavity mirrors separation from the ideal
position. Assuming $\vartheta=0$ and control of the amplitude from
each antenna to $1\%$, the position of the mirrors to $0.1~\mu$m
and taking $1-r=3.6\times 10^{-4}$ (consistent with the Q factor
computed above), we obtain ${\cal R}_{T/S}=(3+9i)\times 10^{-7}$.

\subsubsection{Dipole trap}

We choose a far-detuned dipole trap to contain the atoms for the
duration of the measurement since the perturbations introduced by it
are small and measurable. A variety of different geometries have
been proposed over the years. These include red-detuned traps based
on focused beams, and blue-detuned traps with hollow beams (see
Refs.~\cite{balykin00,friedman02} for reviews of recent work.)

The trap would confine the atoms within $10~\mu$m around the
microwave electric field anti-node and $1$ mm diameter in the radial
dimension. The region of confinement would be smaller than the
microwave wavelength (Lamb-Dicke regime), so Doppler broadening
becomes negligible.

The AC Stark shift($\Delta E$), which produces the restoring force
of the dipole trap, displaces the two hyperfine levels of ground
state in the same direction but not by the same amount. The
differential shift changes the resonant frequency for the
cavity-driven E1 transition used in the anapole moment measurement.
The change in the hyperfine separation for a detuning
($\delta=w-w_e$) larger than the hyperfine splitting
($\Delta_{HFS}$) is approximately equal to $(\Delta_{HFS}/\delta)
\Delta E$~\cite{kaplan02}. The shift reduces considerably using a
blue detuned far off resonance trap (FORT) at 532 nm.

The dipole trap in combination with the cavity field may generate a
multi-photon transition. There are four vectors available for that
transition: ${\bf E1_D}$, ${\bf M1_D}$ the dipole trap electric and
magnetic fields, ${\bf E}$ the microwave electric field and ${\bf
B}$ the static magnetic field. The parity and time reversal
conserving observables created with combinations of the above
vectors that produce a resonant transition (${\bf (E1_D \cdot
E)(M1_D \cdot B)}$, ${\bf (E1_D \cdot B)(M1_D \cdot E)}$, ${\bf
(E1_D \times E) \cdot (M1_D \times B)}$, ${\bf (E1_D \times B) \cdot
(M1_D \times E)}$, ${\bf (E1_D \times M1_D) \cdot (E \times B)}$,
and $i{\bf (E1_D \times E) \cdot M1_D}$), give a negligible
contribution if the trap laser propagates along {\bf B}.

\subsubsection{$M1$ transition}

The dominant transition between the two hyperfine states is a
magnetic dipole M1 transition. The magnetic component of the
microwave field could drive M1 transitions. A microwave magnetic
field polarized along the $x$ axis would have the same signature
as a parity violating signal. The M1 transition amplitude
($A_{M1}$) between the levels of interest is given by

\begin{eqnarray}
A_{M1}/\hbar & & = \langle \overline{f}|(-e/2m_e){\bf (J+S) \cdot
M} |\overline{i}\rangle/\hbar \nonumber \\ & & =7.8\times 10^{6}
\left[\frac{M}{1.6~ {\rm Gauss}} \right]{\rm rad/s}, \label{m1}
\end{eqnarray}
for the maximum expected microwave magnetic field in the
Fabry-Perot cavity. The ratio of the E1 transition (Eq. \ref{e1})
to the M1 transition is $| A_{E1} / A_{M1} | \sim 1 \times
10^{-9}$. The success of the measurement depends on reducing and
understanding this transition. We propose to suppress it in three
ways.

\begin{figure}[h]
\leavevmode \centering
\includegraphics[width=3.1in]{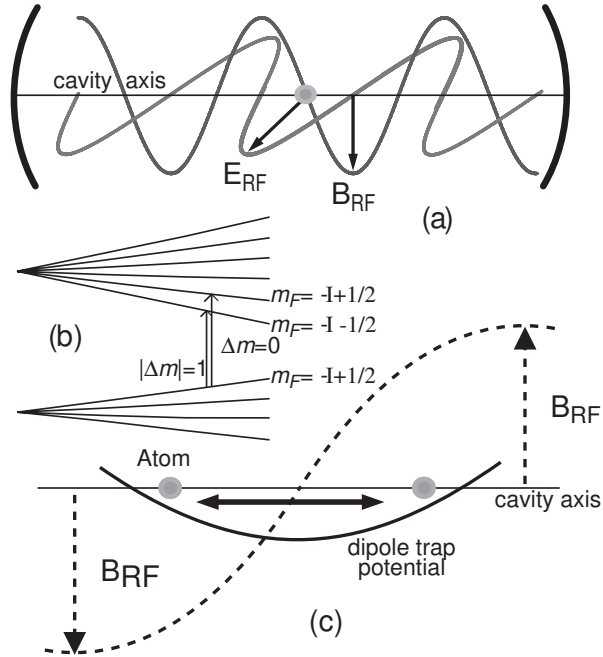}
\caption{Suppression mechanisms of the M1 transition. (a) Trapped
atoms would sit at the magnetic field node, where the magnetic field
is zero. (b) Schematic of the ground hyperfine levels showing a
$|\Delta m|=1$ transition such as the one for the anapole moment and
the $\Delta m=0$ out of resonance transition such as that induced by
the M1 field. The level spacing as well as the spin do not
correspond to any particular atom. (c) Trapped atoms would oscillate
around the microwave magnetic field node and would sample a zero
time-averaged magnetic field. \label{suppression}}
\end{figure}

First (see Fig. \ref{suppression}[a]), we would place the atoms at
the magnetic field node (electric field anti-node) of the microwave
cavity. The magnitude of the microwave magnetic field at the edges
of the atomic trap is reduced by a factor $\aleph=\sin( 2 \pi
d_t/\lambda_m)$, with $d_t=10~\mu$m the length of the trap along the
cavity axis. The reduction factor at 45 GHz is $\aleph=4.8 \times
10^{-3}$.

Second (see Fig. \ref{suppression}[b]), we would direct the
polarization of the M1 field to be along the $z$-axis (Fig.
\ref{setup}). The non-resonant M1 transitions in this case would be
of the type $\Delta m=0$. The static magnetic field ($B_0$) would
split the Zeeman sublevels of the two hyperfine levels, and the
microwave field would be resonant for the $|\Delta m|=1$ E1
transitions (the microwave electric field would be polarized along
the $x$ axis.) The alignment imperfections give a suppression factor
equal to $\sin( \phi ) \sim \phi \sim 10^{-3}$ rad, the angle of the
microwave magnetic field polarization with respect to the $z$ axis.

Third (see Fig. \ref{suppression}(c)), the atoms in the dipole trap
would oscillate around the microwave magnetic field node. An atom
crossing the node would see a microwave magnetic field pointing in
the opposite direction. The change in position effectively would
flip the phase of the magnetic field that the atom sees, and would
reverse the evolution generated by the M1 transition. The dynamical
suppression only takes place if the frequency of oscillation
($\zeta$) of the atoms inside the trap is larger than the Rabi
frequency of the M1 transition and is given by
$(1/\sqrt{N})\Omega_{M1}/\zeta$. The frequency of oscillation along
the cavity axis for the proposed geometry would be $\zeta/2\pi \sim
300$ Hz.

Taken together, the three suppression mechanisms would reduce the
expected M1 transition amplitude to $A_{M1s}/\hbar=1.9\times
10^{-5}$ rad/s for 10$^6$ atoms. This is 500 times smaller than
the amplitude for the E1 transition.

\subsection{Signal to noise ratio}

The magnitude of the signal from Eq. \ref{asymmetry} reaches a
maximum for a Raman transition amplitude of $A_R=(2n+1)\pi/2$ with
$t_R=1$. The measurement of the upper hyperfine state population
collapses the state of each atom into one of the two hyperfine
levels. The collapse distributes the atoms binomially between the
two hyperfine levels and leads to an uncertainty in the measured
excited state fraction called projection noise ${\cal
N}_P$~\cite{itano93}. The projection noise is given by

\begin{equation}
{\cal N}_P=\sqrt{N|c_e|^2(1-|c_e|^2)}. \label{projection}
\end{equation}
The projection noise vanishes when all the atoms are in one of the
hyperfine levels, but in those cases the noise is dominated by other
sources, such as the photon shot noise.

The signal to noise ratio for a projection noise limited
measurement is

\begin{equation}
\frac{\cal S}{{\cal N}_P}=2\frac{A_{E1}t_R}{\hbar} \sqrt{N}.
\label{signaltonoise}
\end{equation}
Taking $A_{E1}$ from Eq. \ref{e1}, $t_R=1$ s, and integrating over
10$^4$ cycles, we would reach a 3$\%$ measurement with only 300
atoms.

The high-efficiency MOT that we developed at Stony Brook, with
production rates around 10$^6$ s$^{-1}$, captures in excess of
10$^5$ francium atoms~\cite{aubin03a}. We expect to trap 10$^6$
atoms after transferring them to an ultra-high vacuum environment.
In this case, Eq. \ref{signaltonoise} predicts a signal to noise
ratio of 20 in 1 s. Higher francium production rates could be
obtained at other facilities, such as ISAC at TRIUMF, where an
actinite target could deliver in excess of 10$^8$ atoms per second
of a single isotope.

While measurements in francium benefit from a large $A_{E1}$,
large atomic samples of other alkalis are easily prepared. We
could obtain the same signal to noise ratio in a cesium sample
with 100 times more atoms and the same strength-driving field.
While the fundamental signal to noise ratio indicates the inherent
trade-offs between different alkali species, technical noise,
specific to the instruments dedicated to the measurement, must
also be considered. For a discussion of technical noise in the
cesium PNC Boulder experiment see Ref. \cite{wood99}.

\section{Noise and Systematic Effects}
\label{technicalnoise}

The measurement of the anapole moment would come from determining
the population transferred from the lower to the upper hyperfine
level by the application of the Raman and microwave fields. Both of
these fields (or any other stray field) are characterized by a field
amplitude, frequency (or detuning), and interaction time. The total
transition amplitude for a common detuning ($\delta$) and
interaction time ($t_R$) is:

\begin{equation}
A = (A_{R1}+A_{E11}+A_1)+ i(A_{R2}+A_{E12}+A_2),
\label{completeomega}
\end{equation}
where $A_{R1,R2}$ are the real and imaginary components of the
Raman amplitude, $A_{E11,E12}$ the corresponding for the E1
transition amplitude and $A_{1,2}$ are the real and imaginary
parts of any other transition present such as an M1 transition.

Table \ref{tablereversals} shows the phase of the transitions for
given field polarizations, with their transformation under magnetic
field reversal assuming all the excitation fields are in phase. We
control the phase difference ($\psi$) between the Raman field and
the cavity E1 field. Varying $\psi$ introduces an additional factor
of $e^{i\psi}$ on the E1 transition amplitude while the Raman
transition remains unchanged. The standing wave M1 field inside of
the cavity is 90$^o$ out of phase with the E1 field, which gives a
factor of $ie^{i\psi}$ for the M1 transition. If instead the M1
field corresponds to a traveling wave, then it is in phase with the
E1 field.

\begin{table}
\leavevmode \centering \caption{Phase ($P=A/|A|$) of the relevant
transition amplitudes for the initial state $F_1=4$, $m_1$ and final
state $F_2=5$, $m_2$ and polarized along the specified axis. For
this table all the fields have the same phase (equal to 0). $P_{Rx}$
represents the Raman transition with one vector along the $y$ axis
and the other along the $z$ axis, such that their cross product
points along the $x$ axis. $\beta$ represents the static magnetic
field reversal together with a sign change on the Zeeman sublevel
$m$.}
\begin{tabular} {lrrrrrrr}
Reversal & m1 & m2 & $P_{E1x}$ & $P_{M1x}$ & $P_{M1y}$ & $P_{Rx}$ & $P_{Ry}$ \\
\hline
Normal   & 0  & -1 & $i$       & 1         & $i$       & $i$      & 1 \\
$\beta$  & 0  & 1  & $-i$      & -1        & $i$       & $-i$     & 1 %
\end{tabular}
\label{tablereversals}
\end{table}

The Raman field would be polarized along the $y$ axis so that
$A_{R1}=A_{Ry}$, and the E1 transition polarized along the $x$ axis
so that $A_{E11}=iA_{E1x}$ (or $\psi=\pi/2$). These two amplitudes
would interfere since both are in phase and only one (the E1)
changes sign under magnetic field reversal as shown in Table
\ref{tablereversals}. Expanding Eq. \ref{excited} for large $A_{Ry}$
compared to the detuning and other amplitudes, we obtain

\begin{equation}
\Xi / N \simeq \sin^2\left(\frac{A_{Ry}t_R}{2\hbar}\right) +\frac12
\sin \left(\frac{A_{Ry}t_R}{\hbar}\right)
\left(\frac{A_{ef}t_R}{\hbar}\right), \label{expansionfirst}
\end{equation}
with,
\begin{equation}
A_{ef}= \left(iA_{Ex}+A_1+ \frac{\hbar^2 \delta^2}{2A_{Ry}}
+\frac{1}{A_{Ry}} (A_{Rx}+A_2)^2 \right). \label{expansion}
\end{equation}
$A_{ef}$ contains the signal ($A_{Ex}$) and noise ($A_1$, $A_2$,
$A_{Rx}$ and $\delta$) terms. We can use this expression to set
limits in the different experimental parameters and identify the
corresponding observable. Expanding the last term in Eq.
\ref{expansionfirst} for small $t_R$ gives

\begin{equation}
\frac{t_R^2}{2\hbar^2}A_{Ry}A_{ef}.
\end{equation}
The first term in $A_{ef}$ is proportional to $iA_{Ry}A_{E1x}$,
which corresponds to the PNC signal $i{\bf (E \times (E}_{R1} {\bf
\times E}_{R2} {\bf ) \cdot B)}$.

The amplitudes of interest are the Raman amplitudes $A_{Rx,Ry}$, the
E1 amplitude $A_{E1x}$, a M1 transition that is in phase with the E1
field $A_{Mix,Miy}$ and an M1 transition that is $\pi/2$ out of
phase with the E1 field $A_{Mox,Moy}$. As an example, if the
standing wave magnetic field inside of the cavity is tilted towards
the $x$ axis it generates an amplitude $A_{Mox}$ since this field is
out of phase with the E1 field. The M1 amplitudes are included in
Eq. \ref{expansion} as $A_1$ or $A_2$ depending on their phase
relation to $A_{Ry}$.

The relevant values for the relative phase ($\psi$) between the E1
and the Raman transition are multiples of $\pi/2$. First we study
the case with $\psi=0,\pi$. This does not correspond to the PNC
measurement since the E1 and Raman transitions are out of phase and
do not interfere. The signal obtained with this configuration is
still helpful in the evaluation of unwanted contributions. We can
rewrite $A_{ef}$ from Eq. \ref{expansion} using Table
\ref{tablereversals} and ignoring the detuning ($\delta$) as

\begin{eqnarray}
A_{ef}& & = \frac{1}{A_{Ry}}\left[ (A_{Mox}^2-A_{Miy}^2-A_{Rx}^2)
\right. \nonumber \\ & & +s(iA_{Ry}A_{Moy}-2iA_{Rx}A_{Mox})
+\beta(-2iA_{Miy}A_{Mox}) \nonumber
\\ & & \left. +s\beta(A_{Ry}A_{Mix}-2A_{Miy}A_{Rx}) \right], \label{psi0}
\end{eqnarray}
with $s=1,-1$ when $\psi=0,\pi$ respectively and $\beta=1,-1$
depending if we have the normal experiment or we apply a magnetic
field reversal. With $\psi=\pi/2,-\pi/2$ instead we get

\begin{eqnarray}
A_{ef} & & =\frac{1}{A_{Ry}}\left[ (A_{Mix}^2-A_{Rx}^2-A_{Moy}^2)
\right. \nonumber \\ & & +s(iA_{Ry}A_{Miy}+2iA_{Rx}A_{Mix})
-\beta(-2iA_{Moy}A_{Mix}) \nonumber \\ & & \left.
+s\beta(2A_{Rx}A_{Moy}-A_{Ry}A_{Mox}+iA_{Ry}A_{E1x}) \right],
\label{psipi}
\end{eqnarray}
where now $s=1,-1$ when $\psi=\pi/2,-\pi/2$ respectively. This
corresponds to the experimental condition for the PNC measurement.
The PNC signal is contained in the last term, and it changes sign
under both reversals. Equations \ref{psi0} and \ref{psipi} show
how reversals can be used to isolate the PNC signal.

We divide the analysis of the different experimental parameters into
three parts: Systematic effects that include terms that mimic the
PNC signal and that are contained in the last parenthesis of Eq.
\ref{psipi}, line broadening mechanisms, which contain all other
terms and that average to zero after an infinite number of cycles,
and calibration errors that modify the value of the extracted
constants on the PNC signal.

\subsection{Line broadening mechanisms}

We start with terms that do not change under both reversals. They
include the detuning term from Eq. \ref{expansion} and all the
terms in Eq. \ref{psipi} except for the last parenthesis. We
present the requirements to achieve a precision of 3\% in the
measurement after $10^4$ repetitions. Each noise amplitude has to
be controlled to $3A_{E1}$

We could reduce the effect of some noise terms by increasing
$A_{Ry}$ (see Eq. \ref{expansion}). We would take $A_{Ry}$ to be
exactly equal to a $(2n+1)\pi/2$ pulse, and include any deviation
from this value into $A_1$. We would control the Raman pulse to
0.025$\%$ in one second with shot noise limited detection. This
would limit the maximum value for the Raman pulse to
$A_{Ry}/\hbar=121$ rad/s or $n=38$. We now proceed to analyze the
spurious terms in Eq. \ref{psipi} that contaminate the signal.

\subsubsection{$\hbar^2 \delta^2 / 2A_{Ry}$}
The detuning can have its origin in poor frequency control on the
microwave or Raman fields, or changes in the external fields that
shift the energy levels. The detuning would have to be controlled to
$\delta=2.7$ rad/s. The required accuracy for the microwave field
frequency is one part in $10^{11}$.

Control of the static magnetic field $B_0$ to a fractional stability
of $5 \times 10^{-5}$ would keep the detuning under control.

The presence of an M1 transition produces an AC shift of the levels.
The value of the maximum shift is $\sim$3 mHz, which is negligible.

The atoms in the trap occupy different vibrational levels.
Transitions between different vibrational levels are suppressed
for a sufficiently far detuned trap. Each vibrational level has
slightly different resonance frequency that leads to broadening of
the signal and loss of coherence.

Coherence times as long as 4.4 s have been measured for atoms in a
blue detuned trap~\cite{davidson95}. The main source for decoherence
was the distribution of Stark shifts felt by the atoms. We expect a
coherence time 16 times smaller in francium than in Ref.
\cite{davidson95} using a laser at 532 nm because of the difference
in hyperfine splitting and detuning. The dephasing grows slowly in
time and can be reversed with the use of an ``echo" technique. The
atoms would spend approximately half of the time in each hyperfine
level with a Raman transition amplitude $A_R=(2n+1)\pi/2$ for large
$n$. It is necessary in that case to keep the coherence for a time
approximately equal to $t_R/n$, with $t_R$ the duration of the
experiment. We would need a coherence time of 26 ms for $n=38$ to
have an interaction time of 1 s. This is below the expected 300 ms
coherence time.

The average differential Stark shift seen by the atoms would be
approximately equal to $kT(\Delta_{HFS}/\delta)/h=6.3$ Hz. The
effect of the time varying detuning generated by the oscillations in
the trap is similar to a steady state detuning of the same
magnitude, and can be compensated by adjusting the microwave
frequency. We must control the power of the trap laser to 7$\%$.

\subsubsection{$A_{Rx}^2/A_{Ry}$}
This term appears due to a bad polarization alignment of the Raman
field. Control of the polarization of the Raman field to one part
in 10$^3$ would be necessary to suppress this term.

\subsubsection{($A_{Mix}^2$, $A_{Moy}^2$, $A_{Rx}A_{Mix}$,
$A_{Moy}A_{Mix})/A_{Ry}$}

These terms are multiplied by a small number and their contribution
becomes negligible. For example, $A_{Rx}A_{Mix}/A_{Ry}$ has the
small factor $A_{Rx}/A_{Ry}$ appearing due to polarization
misalignment in the Raman beams.

\subsubsection{$A_{Ry}A_{Miy}/A_{Ry}= A_{Miy}$}
This is the dominant term that depends on the M1 transition. The M1
field appears due to imperfections in the microwave cavity field
that create a traveling wave component that may be in or out of
phase with the E1 transition.

Eq. \ref{standingwave} gives the amplitude of the traveling wave
expected in our setup. The traveling wave is polarized along the $z$
axis, so we can include the polarization suppression factor of
$10^{-3}$. Combining these two numbers with the amplitude for the M1
transition we get an amplitude of $0.25A_{E1}$ out of phase with the
E1 transition, and an in phase amplitude of $0.75A_{E1}$.

The relative phase between both antennas ($\vartheta$) can be
adjusted by minimizing the M1 contribution when the static
magnetic field ($\bf B$) is tilted slightly. The antennas phase
mismatch contribution remains controlled for $\vartheta<0.01$ rad.

\subsection{Systematic effects}
The systematic effects include the terms in the last parenthesis in
Eq. \ref{psipi}. They change sign under both $s$ and $\beta$
reversals just as the PNC signal. The constraints for these terms
are stronger since they do not average to zero. Their contribution
must be below $0.03A_{E1}$ to reach a 3$\%$ measurement. We proceed
to analyze each one of these terms.

\subsubsection{$A_{Rx}A_{Moy}/A_{Ry}$}
This term appears because of a combination of misalignment of the
Raman field and misalignment of the microwave field or
imperfections in the microwave cavity. It corresponds to the
observable ${\bf M \times (E}_{R1} {\bf \times E}_{R2} {\bf )
\cdot B}$. This term is reduced by the Raman misalignment
($A_{Rx}/A_{Ry}$) and its contribution would become negligible.

\subsubsection{$A_{Ry}A_{Mox}/A_{Ry}=A_{Mox}$}
This term has the same origin as the previous one, but its
contribution is considerably larger since it is not suppressed by
the Raman misalignment. It gives the limiting factor in the
precision of the measurement and its control depends completely on
the suppression mechanisms.

The cavity mirrors may have some birefringence, which generate a
microwave magnetic field $x$-axis component. The microwaves make
roughly 1000 reflections in the cavity. We need a polarization
rotation smaller than $10^{-3}$ rad or a rotation per reflection
smaller than $10^{-6}$ rad to keep the M1 suppression unchanged. The
constraint for a 3$\%$ measurement is 14 times smaller.

The atomic sample would have to be precisely held at the node of
the microwave magnetic field. The maximum displacement we can
tolerate is $3\times 10^{-11}$m for a 3$\%$ measurement.

\subsection{Calibration errors and requirements on theoretical calculations}
The PNC signal (Eq. \ref{asymmetry}) would give directly the
$A_{E1}$ amplitude since the uncertainty in the Raman amplitude is
negligible. $A_{E1}$ is the product of the microwave electric field
and the matrix element. The microwave electric field amplitude has
to be known to 3$\%$. The electric field could be measured by
tilting the magnetic field and inducing an M1 transition. The
extraction of information about the weak interaction from an
experimental measurement requires theoretical
input~\cite{haxton01,ginges04}. The quality of the electronic wave
functions is the most important. The accuracy of the matrix elements
has to be comparable to that of the experiment. The effective
constant of the anapole moment $\kappa_a$ is obtained after
subtracting the other two contributions to $\kappa_{i}$ (Eq.
\ref{kappatotal}). Johnson {\it et al.} show that the other
contributions for the case of Fr amount to a few
percent~\cite{johnson03}. The anapole moment of the even-neutron
isotopes comes only from the unpaired proton, while the odd-neutron
isotopes contain contributions from the unpaired proton and neutron.
A measurement of the anapole moment to better than $10\%$ would give
an initial separation of both contributions~\cite{flambaum97}.

\subsection{Other sources of fluctuations}
The microwave magnetic field would generate transitions to other
levels of the type $\Delta m=0$, which are non-resonant at the
proposed magnetic field (detuning $\sim$ 0.4 GHz). Nevertheless,
these transitions will have to be taken into account in a detailed
analysis of the data.

Stray electric fields produce Stark induced transitions that mimic
the PNC signal. A stray electric field of 13 V/cm in the $z$
direction would generate a transition amplitude equal to the parity
violating signal. Stray fields large enough to be a problem are
unlikely to occur and can be ignored~\cite{demille98}.

Gradients induce higher order multipole transitions, such as an E2
transition. Fortunately, these higher order transitions between the
two hyperfine ground levels are strongly suppressed. Table
\ref{tableconstraints} summarizes the results of the analysis of
noise and systematic effects.

\begin{table}
\leavevmode \centering \caption{Fractional stability required for
a 3$\%$ measurement. The observable associated with each
constraint is also included.}
\begin{tabular} {llrr}
Observable & Constraint & Set value & Stability \\
\hline \\
$A_{Ry}A_{E1}$ & Microwave amplitude & 476 V/cm & $0.03$ \\
$A_{Ry}A_{Ry}$ & Raman amplitude & 121 rad/s & $2.5 \times 10^{-4}$ \\
$(\hbar \delta)^2$ & Microwave frequency & 45 GHz & $10^{-11}$ \\
 & Dipole trap Stark shift & 6.3 Hz & $0.07$ \\
 & DC Magnetic field & 1500 Gauss & $4.7 \times 10^{-5}$ \\
$A_{Rx}A_{Rx}$ & Raman polarization & 0 rad & $10^{-3}$ rad \\
$A_{Ry}A_{Miy}$ & Mirror separation & 13 cm & $7.7 \times 10^{-7}$ \\
 & Antenna power & 57 mW & 0.02 \\
 & Antenna phase & 0 rad & 0.01 rad \\
$A_{Ry}A_{Mox}$ & Mirror birefringence & 0 rad & $1 \times 10^{-4}$ rad \\
 & Trap displacement & 0 m & $3 \times 10^{-11}$ m
\end{tabular}
\label{tableconstraints}
\end{table}

\section{Conclusion}
The anapole moment provides a unique probe of weak hadronic
interactions. In particular it is sensitive to weak long-range meson
exchange interactions, and consequently allows a measurement of weak
neutral currents in the nucleus. This is not the case in high-energy
experiments where the weak contribution must be separated from the
strong and electromagnetic contributions that are much larger. We
have presented the analysis of a proposed measurement strategy of
the nuclear spin dependent part of the PNC interaction, dominated by
the anapole moment. While the proposed measurement method can be
extended to other alkali atoms, a series of measurements in a chain
of francium isotopes allows the separation of the proton and neutron
contributions to the anapole moment.

As noted by Fortson {\it et al.} \cite{fortson90,pollock92} studies
of atomic parity non conservation give information on the nuclear
physics. The nuclear weak interaction at low energies is often
parameterized by a series of coupling constants, either with a meson
exchange formalism, the so called DDH parametrization
\cite{desplanques80}, or more recently with effective field theories
(EFT) \cite{zhu05}. A program of measurements of the anapole moment
in a chain of francium isotopes will contribute significantly to
constraint some of the DDH parameters, which together with the EFT
program will provide a model independent input for theoretical
analysis of low energy weak interaction constants. It is important
to note that the measurement of the anapole moment of an even and an
odd isotope of francium give almost orthogonal bands in the meson
coupling parameter space. This is subject to the assumption that the
anapole moment is carried mainly by the last nucleons
\cite{flambaum97}, but as shown by the measurements of the hyperfine
anomaly \cite{grossman99}, this is a reasonable assumption. These
measurements will significantly contribute to deepen our
understanding of the nuclear structure.

\section*{Acknowledgments}

Work supported by NSF. E. G. acknowledges support from CONACYT.

\end{document}